\RequirePackage{fix-cm}
\documentclass{svjour3}                     
\smartqed  
\usepackage{amsmath,amsfonts,graphicx,bm}
\usepackage{adjustbox,rotating}
\usepackage{multirow}
\usepackage[algo2e]{algorithm2e} 
\usepackage{booktabs}
\usepackage{enumitem}
\usepackage{mathrsfs}
\usepackage{color}
\usepackage{pdflscape,lscape}
\DeclareMathAlphabet\mathpzc{OT1}{pzc}{m}{it}
\let\mathcal=\mathpzc
\def\E{{\mathbb E}}

\numberwithin{equation}{section}

\let\trueiiint=\iiint
\def\iiint{\mathop{\textstyle\trueiiint}\limits}
\def\intinfty{\int\limits_{\!\!-\infty\,\,}^{\,\,\infty\!\!}\kern-0.0em}
\def\iintinfty{\mathop{\int\!\!\int}\limits_{\!\!-\infty\,\,}^{\,\,\infty\!\!}\kern-0.0em}
\def\iiintinfty{\mathop{\int\!\!\int\!\!\int}\limits_{\!\!-\infty\,\,}^{\,\,\infty\!\!}\kern-0.0em}

\let\<=\langle
\let\>=\rangle
\let\^=\hat
\def\~#1{{\-ox{\sf#1}}}

\def\R{{\mathbb R}}

\usepackage{algorithm}
\usepackage{algpseudocode}
\let\-=\mathbf

\def\@#1{{\cal #1}}

\begin{document}

\title{Ensemble Kalman filter based Sequential Monte Carlo Sampler for sequential Bayesian inference\thanks{This work was
supported by the NSFC under grant number 111771289
and by the EPSRC under grant number EP/R018537/1. }} 


\titlerunning{Ensemble Kalman filter based Sequential Monte Carlo Sampler}        

\author{Jiangqi Wu         \and
       Linjie Wen \and
        Peter~L.~Green \and
        Jinglai Li \and
				Simon Maskell 
}

\authorrunning{J. Wu, L. Wen, P. Green, J. Li and S. Maskell} 

\institute{  J. Wu \at
Equipment Reliability Institute,  
Nuclear Power Operations Research Institute, 50 Shiboguan Rd, Shanghai 200120, China.
\and L. Wen \at
              School of Mathematical Sciences,  
Shanghai Jiao Tong University, 800 Dongchuan Rd, Shanghai 200240, China.            
 \and P. L. Green  \at
 School of Engineering, University of Liverpool, Liverpool L69 7XL, UK.\\
\email{plgreen@liverpool.ac.uk}
           \and  J. Li  \at
             School of Mathematics, University of Birmingham, Birmingham B15 2TT, UK.\\
\email{j.li.10@bham.ac.uk}
\and Simon Maskell \at
Department of Electrical Engineering and Electronics,
University of Liverpool, Liverpool L69 7XL, UK.\\
\email{S.Maskell@liverpool.ac.uk}
}

\date{Received: date / Accepted: date}

\maketitle

\begin{abstract}
Many real-world problems require one to estimate parameters of interest, in a Bayesian framework, from data that are 
collected sequentially in time. Conventional methods for sampling from posterior distributions, such as {Markov Chain Monte Carlo} can not efficiently address such problems as they do not take 
advantage of the data's sequential structure. 
To this end, sequential methods which seek
to update the posterior distribution whenever a new collection of data become available are often used to solve these types of problems.   
Two popular choices of sequential method are the Ensemble Kalman filter (EnKF) and the sequential Monte Carlo sampler (SMCS).
While EnKF only computes a Gaussian approximation of the posterior distribution,  SMCS can draw samples directly from the posterior. Its performance, however, 
depends critically upon the kernels that are used. In this work, we present a method that constructs
the kernels of SMCS using an EnKF formulation, and we demonstrate the performance of the method with numerical examples. 

\end{abstract}



\section{Introduction}
Bayesian inference is a popular method for estimating unknown parameters from data,
largely due to its ability to quantify uncertainty in the estimation results~\cite{gelman2014bayesian}. 
In the current work we consider a special class of Bayesian inference problems where data have to be collected in a sequential manner. 
A typical example of this type of problem is the estimation of parameters,
such as the initial states or the equation coefficients, in a dynamical system from observations related to the state vector at discrete times. 
Such problems arise from many real-world applications, ranging from weather prediction~\cite{annan2004efficient} to biochemical networks~\cite{golightly2011bayesian}.
It should be emphasized that, unlike many data assimilation problems that seek
to estimate the time-dependent states in dynamical systems, the parameters that we want to estimate 
here are assumed  not to vary in time. To distinguish the two types of problems, we refer to the former as \emph{state estimation} problems
and the latter as \emph{parameter estimation}. 
We should also note that in this work we focus on methods which use samples to represent the posterior distribution, 
and that approximation based methods, such as the Variational Bayes~\cite{beal2003variational} and the Expectation Propagation~\cite{minka2001appBI}
will not be discussed here. 
Conventional sampling methods, such as Markov Chain Monte Carlo (MCMC) simulations~\cite{gilks1995markov}, use all the data in a single batch,
are unable to take advantage of the sequential structure of the problems. 
On the other hand, sequential methods utilize the sequential structure of the problem and update the posterior whenever a new collection of data become available,
which makes them particularly convenient and efficient for sequential inference problems.  

A popular sequential method for parameter estimation is the Ensemble Kalman filtering (EnKF) algorithm, 
which was initially developed to address the dynamical state estimation problems~\cite{evensen2009data}. 
The EnKF  method was extended to estimate parameters 
in many practical problems, e.g., \cite{annan2004efficient,annan2005parameter},
and more recently it was generically formulated as a derivative-free optimization
based parameter estimation method in \cite{iglesias2013ensemble}.
The EnKF method for parameter estimation was further developed and analyzed in 
\cite{arnold2014parameter,iglesias2016regularizing,schillings2017analysis}, etc.
The basic idea of the EnKF method for parameter estimation is to construct an artificial dynamical system,
turning the parameters of interest into the states of the constructed dynamical system, 
 before applying the standard EnKF procedure to estimate the states of the system. 
 A major limitation of the EnKF method is that, 
 just like the original version for dynamical state estimation, it can only compute a Gaussian approximation 
 of the posterior distribution, and thus methods directly sampling the posterior distribution are certainly desirable. 
To this end, the Sequential Monte Carlo sampler (SMCS) method~\cite{del2006sequential}, can draw samples directly from the posterior distribution.
The SMCS algorithm is a generalisation of the particle filter~\cite{arulampalam2002tutorial,doucet2009tutorial} for dynamic state estimation, 
generating weighted samples from the posterior distribution. 
Since the SMCS algorithm was proposed in \cite{del2006sequential}, considerable improvements and extensions 
of the method have been proposed, such as, \cite{fearnhead2013adaptive,beskos2017multilevel,heng2020controlled,everitt2020sequential},
and more information on the developments of the SMCS methods can be found in the recent reviews~\cite{dai2020invitation,chopin2020introduction}.
We also  note here that there are other parameter estimation schemes also based on particle filtering, e.g., \cite{gilks2001following,chopin2002sequential},
and the differences and connections between SMCS and these schemes are discussed in \cite{del2006sequential}.  
The SMCS  method makes no assumption
or approximation of the posterior distribution,
and can directly draw (weighted) samples from any posterior. As will be discussed later, a key issue in the implementation of  SMCS is to choose suitable forward and backward kernels, as the performance of SMCS depends critically on such choices. 
As has been shown in \cite{del2006sequential}, the optimal forward and backward kernels exist in principle, 
but designing effective kernels for specific problems is nevertheless a highly challenging task.  
In dynamic state estimation problems, often the EnKF approximation is used as the proposal distribution 
in the particle filtering algorithm~\cite{papadakis2010data,wen2018defensive}, especially for problems 
in which the posteriors are only modestly non-Gaussian.  
Building upon similar ideas,  we propose in this work to construct the kernels in SMCS by using an EnKF 
framework. Specifically, the forward kernel is obtained directly from an EnKF approximation,
and the backward kernel is derived by making a Gaussian approximation of the optimal backward kernel. 

The remaining work is organized as follows. In Section~\ref{sec:setup} we present the generic setup of the sequential inference problems that we consider in this work. 
In Sections~\ref{sec:smcs} and \ref{sec:enkf} we respectively review the SMCS and the EnKF methods for solving sequential inference problems. 
In Section~\ref{sec:enkf-smcs} we present the proposed EnkF-SMCS method and in Section~\ref{sec:examples} we provide several numerical 
examples to illustrate the performance of the proposed method. 
Finally Section~\ref{sec:conclusions} offers some concluding remarks.

\section{Problem setup} \label{sec:setup}
We consider a sequential inference problem formulated as follows. 
Suppose that we want to estimate the parameter $x\in \R^{n_x}$ from data $y_1, ..., y_t, ..., y_T$ which become available sequentially in time. 
In particular the data $y_t\in \R^{n_y}$ is related to the parameter of interest $x$ via the follow model, 
\[ y_t = G_t(x) + \eta_t, \quad t=1...T,\]
where each $G_t(\cdot)$ is a mapping from $\R^{n_x}$ to $\R^{n_y}$, and the observation noise $\eta_t \sim \@N(0, R_t)$.
It follows that the likelihood function can be written as,  
\begin{equation} \pi(y_t|x) = \@N(G_t(x),R_t),\quad t=1...T.\label{e:lh}\end{equation}
It is important to note here that the restriction that the error model has to be additive Gaussian
as is in Eq.~\eqref{e:lh} is due to the use of EnKF.  
While noting that relaxing such a restriction is possible, we emphasize here that additive Gaussian assumption noise is reasonable for a wide range of practical problems. 
We can now write the posterior distribution in a sequential form:
\begin{equation}
\pi_t(x)=\pi(x|y_1,...y_t) \propto \pi_0(x) \prod_{i=1}^t\pi(y_i|x), \label{e:pt}
\end{equation} 
where $\pi_0(x)$ is the prior distribution of $x$, 
and our goal is to draw samples from $\pi_t$ for any $0<t\leq T$. 


 The posterior in Eq.~\eqref{e:pt} is essential in a data tempering formulation, 
and as is pointed out in \cite{fearnhead2013adaptive,zhou2016toward}, such problems pose challenges for usual MCMC methods especially 
when the amount of data is large, as they
cannot conveniently  exploit the sequential structure of the problem. 
In what follows, we first discuss two sequential methods for this type of problems: the EnKF and the SMCS algorithms,
and we then propose a scheme to combine these two methods.

\section{Sequential Monte Carlo Sampler}\label{sec:smcs}
We first give a brief introduction to the SMCS method for sampling 
the posterior distribution $\pi_t(x)$, following \cite{del2006sequential}. 
The key idea of SMCS is to construct a joint distribution $\pi(x_1,...,x_t)$,
the marginal of which is equal to the target distribution $\pi_t(\cdot)$. 
Note here that $\pi(x_1,...,x_t)$ needs only to be known up to a normalization constant. 
One then applies the sequential importance sampling algorithm~\cite{arulampalam2002tutorial,doucet2009tutorial} to draw weighted samples 
from  $\pi(x_1,...,x_t)$, which after being marginalized over $x_1,...,x_{t-1}$, yields 
samples from $\pi_t(\cdot)$. 

Next we describe SMCS in a recursive formulation where, given an arbitrary conditional distribution $L_{t-1}(x_{t-1}|x_t)$,
we can construct a joint distribution of $x_{t-1}$ and $x_{t}$ in the form of, 
\begin{equation}
p_t(x_{t-1},x_t)=\pi_t(x_t) L_{t-1}(x_{t-1}|x_t).
\end{equation} 
such that the marginal distribution of $p_t(x_{t-1},x_t)$ over $x_{t-1}$ is $\pi_t(x_t)$.
Now, given a marginal distribution $q_{t-1}(x_{t-1})$
and a conditional distribution $K_{t}(x_t|x_{t-1})$, we can construct an importance sampling (IS) distribution for $p_t(x_{t-1},x_t)$ in the form of
\begin{equation}
q_t(x_{t-1},x_t) = q_{t-1}(x_{t-1})K_{t}(x_t|x_{t-1}). \label{e:isq}
\end{equation}
It is important to note here that a key requirement of the IS distribution $q_t(x_{t-1},x_t)$ is that we can directly 
draw samples from it.
We let $\{x^m_{t-1:t}\}_{m=1}^M$ be an ensemble drawn from $q_t(x_{t-1},x_t)$, and note that the weighted ensemble
 $\{(x^m_{t-1:t},w_{t}^m)\}_{m=1}^M$ follows the distribution $p_t(x_{t-1:t})$, where the weights are computed according to
\begin{subequations}\label{e:isw}
\begin{eqnarray}
w_t(x_{t-1:t}) &=& \frac{p_t(x_{t-1},x_t)}{q_t(x_{t-1},x_t)}
 = \frac{\pi_t(x_t) L_{t-1}(x_{t-1}|x_t)}{q_{t-1}(x_{t-1})K_{t}(x_t|x_{t-1})} \notag\\
&=& w_{t-1}(x_{t-1}) \alpha_t(x_{t-1},x_t),
\end{eqnarray}
where 
\begin{equation}
w_{t-1}(x_{t-1}) = \frac{\pi_{t-1}(x_{t-1})}{q_{t-1}(x_{t-1})},\quad
\alpha_t(x_{t-1},x_t)=\frac{\pi_t(x_t) L_{t-1}(x_{t-1}|x_t)}{\pi_{t-1}(x_{t-1})K_{t}(x_t|x_{t-1})}.
\end{equation}
\end{subequations}
As can be seen here, once the two conditional distributions  $K_t$ and $L_{t-1}$ (respectively referred to  
as the forward and backward kernels in the rest of the paper) are chosen, we can draw 
samples from Eq.~\eqref{e:isq} and compute the associated weights from Eq.~\eqref{e:isw},
obtaining weighted samples from $p_t(x_{t-1},x_t)$ as well as its marginal $\pi_t(x_t)$.
The SMCS essentially conducts this procedure in the following sequential manner:
\begin{enumerate}
\item  let $t=0$, draw an ensemble $\{x^m_{0}\}_{m=1}^M$ from $q_0(x_0)$,
and compute $w^m_0=\pi_0(x^m_0)/q_0(x_0^m)$ for $m=1...M$;
\item let $t=t+1$; \label{st:t+1}
\item draw $x^m_{t}$ from $K(\cdot|x^m_{t-1})$ for each $m=1...M$;
\item compute $w^{m}_{t}$ using Eq.~\eqref{e:isw};
\item  return to step~\ref{st:t+1} if $t<T$.
\end{enumerate}


Note here that, a resampling step is often used in SMCS algorithm to alleviate the ``sample degeneracy'' issue \cite{del2006sequential}. 
The resampling techniques are well documented in the PF literature, e.g., ~\cite{arulampalam2002tutorial,doucet2009tutorial}, and so are not discussed here. \\

As can be seen from the discussion above, to use SMCS one must choose the two kernels. 
In principle, optimal choices of these kernels are available. 
For example, it is known that once $K_t(x_{t}|x_{t-1})$ is provided, one can derive that the optimal choice of $L_{t-1}(x_{t-1}|x_t)$ 
is~\cite{del2006sequential}:
\begin{eqnarray}
L^\mathrm{opt}_{t-1}(x_{t-1}|x_t) &=& \frac{q_{t-1}(x_{t-1}) K_t(x_t|x_{t-1})}{q_t(x_t)}   \notag\\
&=& {  \frac{q_{t-1}(x_{t-1}) K_t(x_t|x_{t-1})}{{\int q_{t-1}(x_{t-1}) K_t(x_t|x_{t-1}) dx_{t-1}}}, }
\label{e:optL}
\end{eqnarray}
where the optimality is in the sense of yielding the minimal estimator variance. We also note that use of the optimal L-kernel allows the weights to be written as 
\begin{equation}
w_t(x_{t-1:t}) = \frac{\pi_t(x_t)}{q_t(x_t)}. \label{e:optw}
\end{equation}
Moreover, we can see here that if we can choose $K_t$ such that $q_t= \pi_t$, then the weight function is always unity,
which means that we now sample directly from the target distribution (the ideal case). While obtaining such an ideal $K_t$ is usually not possible in practice, 
it nevertheless provides useful guideline regarding choice of the forward kernel $K_t$ i.e.
it should be chosen such that the resulting $q_t$ is close to $\pi_t$. 
For example,  it is proposed in \cite{del2006sequential} to use the MCMC moves as the forward kernel.
A main limitation of the MCMC kernel is that it typically requires a number of MCMC moves
to propose a ``good'' particle, and since each MCMC move involves an evaluation of the underlying mathematical model, $G_t$,
the total computational cost can be high when $G_t$ is computationally intensive.

In this work we consider an alternative to the use of MCMC kernels. 
Specifically we propose to choose $K_t$ of the form
\begin{equation}
K_t(\cdot|x_{t-1}) = \@N(\cdot|T_{t}(x_{t-1}),\Sigma^K_t),
\end{equation}
i.e., a Gaussian distribution with mean $T_t(x_{t-1})$ and variance $\Sigma^K_t$,
where $T_t(\cdot)$ is a $\R^{n_x}\rightarrow \R^{n_x}$ transformation. 
{We shall compute $T_t$ and $\Sigma^K_t$ (or equivalently  the forward kernel $K_t$) using 
the EnKF method.}

\section{Ensemble Kalman Filter}\label{sec:enkf}
In this section we give a brief overview of the EnKF
parameter estimation method proposed in \cite{iglesias2013ensemble},
which essentially  aims to compute a Gaussian approximation of $\pi_t(x_t)$ in each time step $t$. 
To formulate the problem in an EnKF framework, we first construct an \emph{artificial dynamical system} denoted by $F_t$; at any time $t$, we have the
 states $u_t=[x_t,z_t]^T$ where $z_t=G_t(x_t)$,
and the dynamical model, 
\begin{equation}
u_t= F_t(u_{t-1}),\quad x_t= x_{t-1},\quad {z}_t = G_t(x_{t}).\label{e:prop}
\end{equation}
The data is associated to the states through $y_t = z_t+\eta_t$,
 or equivalently 
\[y_t = H u_t +\eta_t = [\it0_{n_y\times n_x}, I_{n_y\times n_y}] u_t+\eta_t,\]
where $I_{n_y\times n_y}$ is a $n_y\times n_y$ identity matrix and $\it0_{n_y\times n_x}$ is a $n_y\times n_x$ zero matrix.
We emphasize here that once we have the posterior distribution $\pi(u_t|y_{1:t})$, we can obtain the posterior 
$\pi_t(x_t)=\pi(x_t|y_{1:t})$ by marginalizing $\pi(u_t|y_{1:t})$ over $z_t$.  

Now let us see how the EnKF proceeds to compute a Gaussian approximation of the posterior distribution $\pi(u_t|y_{1:t})$. 
At time $t$, suppose that the prior $\pi(u_{t}|y_{1:t-1})$ can be approximated by a Gaussian distribution with 
 mean $\tilde{\mu}_{t}$ 
and covariance $\tilde{C}_{t}$. 
It follows that the posterior distribution $\pi(u_{t}|y_{1:t})$ is also  Gaussian and its mean and covariance can be obtained analytically: 
\begin{equation}
{\mu}_t = \tilde{\mu}_t +Q_t(y_t-H\tilde{\mu}_t), \quad {C}_t = (I-Q_tH)\tilde{C}_t , \label{e:postparams}
\end{equation}
where $I$ is the identity matrix and
\begin{equation}
Q_t =\tilde{C}_t H^T(H\tilde{C}_t H^T+R_t)^{-1}\label{e:gain}
\end{equation}
 is the so-called Kalman gain matrix. 

In the EnKF method, one avoids computing the mean and the covariance directly in each step. 
Instead, both the prior and the posterior distributions are represented with a set of samples. 
Suppose that
 at time $t-1$, we have an ensemble of particles $\{u_{t-1}^m\}_{m=1}^M$ drawn according to the posterior distribution $\pi(u_{t-1}|y_{1:t-1})$,   we can propagate the 
 particles via the dynamical model~\eqref{e:prop}: 
 \begin{equation}
 \tilde{u}_t^m = F_t(u_{t-1}^m),
 \end{equation}
 for $m=1...M$, obtaining 
an assemble $\{\tilde{u}_t^m\}_{m=1}^M$
following the prior $\pi(u_{t}|y_{1:t-1})$.
We can compute a Gaussian approximation,  $\@N(u_t | \tilde{\mu}_t, \tilde{C}_t)$, of $\pi(u_{t}|y_{1:t-1})$, 
where the mean and the covariance of $\pi(u_{t}|y_{1:t-1})$ are estimated from the samples:
\begin{equation}
\tilde{\mu}_t = \frac1M\sum_{m=1}^M \tilde{u}_{t}^m, \quad \tilde{C}_t=\frac1{M-1}\sum_{m=1}^M(\tilde{u}_t^m-\tilde{\mu}_t )(\tilde{u}_t^m-\tilde{\mu}_t )^T. \label{e:priorparams}
\end{equation}
Once $\tilde{\mu}_t$ and $\tilde{C}_t$ are obtained, we then can compute $\mu_t$ and $C_t$ directly from 
Eq.~\eqref{e:postparams}, and by design, 
the posterior distribution $\pi(u_t|y_{1:t})$ is approximated by $\@N({\mu}_t, {C}_t)$.
Moreover it can be verified that the samples 
\begin{equation}
{u}_{t}^m =\tilde{u}_t^m +Q_t(y_t-(H\tilde{u}_{t}^m-\eta^m_t)), \quad\eta_t^m\sim \@N(0,R_t), \quad m=1...M, 
\label{e:update}
\end{equation}
with $Q_t$ computed by Eq.~\eqref{e:gain}, follow the distribution $\@N({\mu}_t,{C}_t)$.
That is, $\{u_{t}^m\}_{m=1}^M$ are the approximate ensemble of  $\pi(u_t|y_{1:t})$,
and consequently the associated $\{x_{t}^m\}_{m=1}^M$ approximately follows distribution $\pi_t(x_t)= \pi(x_t|y_{1:t})$.

\section{EnKF-SMCS} \label{sec:enkf-smcs}

Now we shall discuss how to use the EnKF scheme to construct the forward kernel $K_t$ for SMCS. 
First recall that $u_t=[x_t,z_t]^T$, $H =[\it0_{n_y\times n_x}, I_{n_y\times n_y}]$ and the propagation model $x_t=x_{t-1}$,
and we can derive from Eq.~\eqref{e:update} that, 
\begin{subequations}
\begin{align} \label{e:updatex}
x_t= x_{t-1}+Q_t^x (y_t-G_t(x_{t-1})) + Q^x_t\eta_t+\eta'_t,\\
\eta_t\sim \@N(0,R_t),\,  
\eta'_t\sim \@N(0,\delta^2\Sigma^q_{t-1}), 
 \end{align}
 \end{subequations}
where $\delta$ is a small constant, $\Sigma^q_{t-1}$ is the covariance of $q_{t-1}$ (the evaluation
of $\Sigma^q_{t-1}$ is provided in Eq.~\eqref{e:estqt-1}),
and {   $Q^x_t$ is a submatrix of $Q_t$ formed taking the first $n_x$ rows 
and the first $n_y$ columns of $Q_t$,
denoted as $Q^x_t = Q_t[1:n_x,1:n_y]$.}
Eq.~\eqref{e:updatex} can also be written as a conditional distribution: 
\begin{subequations}\label{e:Kt}
\begin{equation}
K_t(\cdot|x_{t-1}) = \@N(\cdot | T_t(x_{t-1}), \Sigma^K_t),
\end{equation}
where
 \begin{equation}T_t(x_{t-1}) = x_{t-1}+Q^x_t (y_t-G_t(x_{t-1}))\quad
\mbox{and}\quad\Sigma^K_t=Q^x_tR_t(Q^x_t)^T+\delta^2\Sigma^q_{t-1}.\end{equation}
\end{subequations}
Note that the purpose of introducing the small noise term, $\eta'_{t}$, in Eq.~\eqref{e:updatex} is to ensure that 
$\Sigma^K_t$ is strictly positive definite and so $K_t$ is a valid Gaussian conditional distribution.
In all the numerical implementations performed in this work, $\delta$ is set to be $10^{-4}$. 
According to the discussion in Section~\ref{sec:enkf},
we have, if $q_{t-1}$ is a good approximation to $\pi_{t-1}$ 
\begin{equation}
q_t(x_{t}) = \int K_t(x_t|x_{t-1}) q_{t-1}(x_{t-1}) d x_{t-1} \approx \pi_t(x_t). \label{e:qt}
\end{equation}
That is,   
 Eq.~\eqref{e:Kt} provides a good forward Kernel for the SMC sampler. 
 It should be noted here that since $T_t$ is a nonlinear transform, 
  in general, we can not derive the analytical expression for $q_t$ and as a result, we can not use the optimal backward kernel given
  in Eq.~\eqref{e:optL}. 
%
Nonetheless,  we can use a sub-optimal backward kernel: 
\begin{equation}
\hat{L}_{t-1}(x_{t-1}|x_t) = \frac{\hat{q}_{t-1}(x_{t-1}) \hat{K}_t(x_t|x_{t-1})}{\int \hat{q}_{t-1}(x_{t-1}) \hat{K}_t(x_t|x_{t-1}) dx_{t-1}}, \label{e:suboptL}
\end{equation}
where $\hat{q}_{t-1}$ is  the Gaussian approximation of $q_{t-1}$,
 and  $\hat{K}_t$ is an approximation of ${K}_t$.
 Next we need to determine $\hat{q}_{t-1}$ and $\hat{K}_t$.
Here $\hat{q}_{t-1}$ can be estimated from the ensemble $\{x^m_{t-1}\}_{m=1}^M$:
\begin{subequations} \label{e:qhat}
  \begin{align}
  \hat{q}_{t-1}(\cdot) &=\@N(\cdot | \xi_{t-1},{\Sigma}^q_{t-1}),\\
  \xi_{t-1} &= \frac{1}{M}\sum_{m=1}^M x_{t-1}^m, \quad
 {\Sigma}_{t-1}^q =\frac1{M-1}\sum_{m=1}^M({x}_{t-1}^m-\xi_{t-1} )(x_{t-1}^m-\xi_{t-1} )^T. \label{e:estqt-1}
 \end{align}
 \end{subequations}
 Now recall that the issue with the optimal backward kernel $L^\mathrm{opt}_{t-1}$ is 
 that the transform $T_t$ inside the forward kernel $K_t$ is nonlinear, and as a result $q_t$ can not be computed analytically. 
 Here to obtain $\hat{L}_{t-1}$ in Eq.~\eqref{e:suboptL} explicitly, we take 
 \begin{equation}
 \hat{K}_t(\cdot|x_{t-1}) = \@N(\cdot | x_{t-1}+Q^x_t (y_t-\bar{y}_t), \Sigma_t^K),\quad\mathrm{with} \quad \bar{y}_t=\E_{x_{t-1}|y_{1:t-1}}[G_t(x_{t-1})],\label{e:hatKt}
 \end{equation}
 and in practice $\bar{y}$ is evaluated from the particles, i.e.,
 \begin{equation}
 \bar{y} = \frac1M\sum_{m=1}^M G_t(x_{t-1}^m).
 \end{equation}
It follows that the backward kernel $\hat{L}_{t-1}$, given by Eq.~\eqref{e:suboptL}, is also Gaussian
and is given by  
\begin{subequations}\label{e:suboptLnormal}
\begin{equation}
\hat{L}_{t-1}(\cdot|x_t) = \@N(\cdot | T^L_{t-1}(x_t),\Sigma^L_{t-1}),
\end{equation} 
where 
\begin{multline}
T_{t-1}^L(x_{t}) =(I-\Sigma_{t}^K(\Sigma^K_t+\Sigma^q_{t-1})^{-1})(x_t-Q_t^x(y_t-\bar{y}_t))
\\+(I-\Sigma_{t-1}^q(\Sigma^q_{t-1}+\Sigma^K_{t})^{-1})\xi_{t-1},
\end{multline}
and
\begin{equation}
 \Sigma^L_t= \Sigma^q_{t-1}-\Sigma^q_{t-1}(\Sigma_{t-1}^q+\Sigma^K_t)^{-1}\Sigma_{t-1}^q. \label{e:alpha1}
\end{equation}
\end{subequations}
It follows that the resulting incremental weight function is 
\begin{equation}
\alpha_t(x_{t-1},x_t)=\frac{\pi_t(x_t) \hat{L}_{t-1}(x_{t-1}|x_t)}{\pi_{t-1}(x_{t-1})K_{t}(x_t|x_{t-1})}.\label{e:alpha1}
\end{equation}
Now using the   ingredients presented above,  we summarize the  EnKF-SMCS scheme
in Algorithm~\ref{alg:enkf-smcs}.

\begin{algorithm}
 Initialization: draw sample $\{x_0^m\}_{m=1}^M$ from distribution $q_0(x_0)$;
  compute the weights $w^m_0=\pi_0(x_0^m)/q_0(x_0^m)$ for $m=1...M$ and renormalize
  $\{w_0^m\}_{m=1}^M$ so that $\sum_{m=1}^M w^m_0=1$;\;\\
 \For{$t=1$ \KwTo $T$}{
 estimate $\xi_{t-1}$ and ${\Sigma}^q_{t-1}$ from the ensemble $\{(x_{t-1}^m,w^m_{t-1})\}_{m=1}^M$ using Eq.~\eqref{e:estqt-1};\; \\ 
let $\tilde{u}_t^m = [x_{t-1}^m, G_t(x_{t-1})^m]^T$ for $m=1...M$;\;\\
 evaluate $\tilde{\mu}_t$ and $\tilde{C}_t$ with Eq.~\eqref{e:priorparams}, and 
 compute $Q_t$ with Eq.~\eqref{e:gain};\;\\
 draw $x_{t}^m \sim K_t(x_t|x_{t-1}^m)$ for $m=1...M$ with $K_t$ given by Eq.~\eqref{e:Kt};\;\\
 compute $\hat{L}_{t-1}$ from Eq.~\eqref{e:suboptLnormal};\;\\
update the weights:
 \[w_{t}^m
= w_{t-1}^m\frac{\pi_t(x^m_t) \hat{L}_{t-1}(x^m_{t-1}|x^m_t)}{\pi_{t-1}(x^m_{t-1})K_{t}(x^m_t|x^m_{t-1})}
\]
and renormalize $\{w_t^m\}_{m=1}^M$ so that $\sum_{m=1}^M w^m_t=1$;\;\\
resample if needed. 
  }
	\caption{The EnKF-SMCS algorithm}\label{alg:enkf-smcs}
\end{algorithm}

{ It is important to note that a key challenge is yet to be addressed in Algorithm~\ref{alg:enkf-smcs}, 
namely the computational cost of computing the particle weight.
First recall that the main computational cost 
arises from the evaluation of the forward model $G_t$, and therefore, the total computational cost can be 
approximately measured by the number of evaluations of $G_t$. 
We can see from Eq.~\eqref{e:alpha1} that when updating the particle weight, we need to compute $\pi_t(x_t)$, 
which involves the evaluation of the forward model from $G_1$ to $G_t$.
This operation is required at each time step, and 
as a result  the number of the model evaluations is at the order of $O(T^2)$ for each particle.
Therefore, the total computational cost can be prohibitive if $T$ is large. }
We propose a method to tackle the issue, which is based on the following two observations.
{First, here we mainly consider the sequential inference problems where one is primarily interested in the posterior distribution at the final step where all data are incorporated
 (we appreciate that there are problems where  all the intermediate posteriors are of interest
 and in this case the method proposed here does not apply).}
{Second, in many practical problems, after some number of observations, the posteriors may not vary substantially in several consecutive steps.}
It therefore may not be necessary to \emph{exactly} compute the posterior distribution at each time step
and, as a result, we only need to sample the posterior distribution in a relatively small number of selected steps. 
Based on this idea, we propose the following scheme in each time step to reduce the computational cost:
we first compute an approximate weight for each particle,  and then assess that if some prescribed conditions (based on the approximate weights) are satisfied. If such conditions are satisfied,  we evaluate the actual weights of the particles. 
To implement this scheme, we have to address the following issues:
\begin{itemize}
\item First we need a method to compute the approximate weight, which should be  much easier to compute than the exact weight.
Recall that in Eq.~\eqref{e:alpha1} one has to evaluate $\pi_t(\-x_t)/\pi_{t-1}(\-x_{t-1})$ which involves 
computing the forward models from $G_1(\-x_t)$ all the way to $G_t(\-x_t)$,
and so the computational cost is high. To reduce the computational cost, we propose the following approximate method to evaluate Eq.~\eqref{e:alpha1}. Namely we first write  $\pi_t(\-x_t)/\pi_{t-1}(\-x_{t-1})$ as, 
\[
\frac{\pi_t(\-x_t)}{\pi_{t-1}(\-x_{t-1})} = \frac{\pi_{t-1}(\-x_t)}{\pi_{t-1}(\-x_{t-1})} \pi(y_{t}|\-x_t),
\]
and naturally we can approximate $\pi_{t-1}$ with $q_{t-1}$, yielding,
\[\frac{\pi_t(\-x_t)}{\pi_{t-1}(\-x_{t-1})}
\approx \frac{{q}_{t-1}(\-x_t)}{{q}_{t-1}(\-x_{t-1})} \pi(y_{t}|\-x_t).\]
{Though $q_{t-1}$ is formally given by Eq.~\eqref{e:qt}, it is not  computationally tractable.}
Thus we make another approximation, replacing $q_{t-1}$ with $\hat{q}_{t-1}$, 
where {$\hat{q}_{t-1}$} is the Gaussian approximation of $q_{t-1}$ given by Eqs.~\eqref{e:qhat},
and as a result, we obtain
\begin{equation}
\alpha_t(x_{t-1},x_t)\approx\frac{\hat{q}_{t-1}(x_t)\pi(y_{t}|\-x_t) \hat{L}_{t-1}(x_{t-1}|x_t)}{\hat{q}_{t-1}(x_{t-1})K_{t}(x_t|x_{t-1})},\label{e:alpha12}
\end{equation}
which is used to compute the approximate weights. 

\item Second we need to prescribe the conditions for triggering the computation of the actual weights. 
Following \cite{green2017estimating}, we use the Effective Sample Size (ESS) \cite{doucet2009tutorial} (based on the approximate weights) as the main indicator for computing 
the actual weights.
Namely if the ESS calculated with the approximate weights is smaller than a threshold value, the 
actual weights are computed.
Moreover we also have two additional  conditions that can also trigger the computation of the actual weights: 1) if the actual weights have not been computed for  a given number of steps;
2) if the inference reaches the final step, i.e., $t=T$. 
We refer to such a step as a weight refinement. 
\item Finally we shall discuss how to compute the actual weight $w_t$. 
It should be noted here that the recursive formulas~\eqref{e:isw} can not be used here 
since the actual value of $w_{t-1}$ is not available. 
However, letting $t_0$ be the preceding step where the actual weights are computed, and it can be shown that 
\begin{equation}
w_t = w_{t_0} \frac{\pi_t(x_t)}{\pi_{t_0}(x_{t_0})}\prod_{i=t_0}^{t-1} \frac{\hat{L}_i(x_i|x_{i+1})}{K_{i+1}(x_{i+1}|x_{i})},
\end{equation}
which is used to calculate the actual weights of the particles. 
\end{itemize}
 We refer to this modified scheme as EnKF-SMCS with weight refinement (EnKF-SMCS-WR), 
the complete procedure of which is described in Algorithm~\ref{alg:enkf-smcs2}.
Note here that in both EnKF-SMCS algorithms, a resampling step is needed.   
{ Finally we can see that, in EnKF-SMCS-WR, the number of forward model evaluations can potentially 
be significantly reduced, and the actual number of evaluations depends on how frequently the weight refinement is triggered.}

\begin{algorithm}
 Initialization: draw sample $\{x_0^m\}_{m=1}^M$ from distribution $q_0(x_0)$;
  compute the weights $w^m_0=\pi_0(x_0^m)/q_0(x_0^m)$ for $m=1...M$ and renormalize
  $\{w_0^m\}_{m=1}^M$ so that $\sum_{m=1}^M w^m_0=1$;\;\\
  let $t_0=0$;\\
 \For{$t=1$ \KwTo $T$}{
 estimate $\xi_{t-1}$ and ${\Sigma}^q_{t-1}$ from the ensemble $\{(x_{t-1}^m,w^m_{t-1})\}_{m=1}^M$ using Eq.~\eqref{e:estqt-1};\; \\ 
let $\tilde{u}_t^m = [x_{t-1}^m, G_t(x_{t-1})^m]^T$ for $m=1...M$;\;\\
 evaluate $\tilde{\mu}_t$ and $\tilde{C}_t$ with Eq.~\eqref{e:priorparams}, and 
 compute $Q_t$ with Eq.~\eqref{e:gain};\;\\
 draw $x_{t}^m \sim K_t(x_t|x_{t-1}^m)$ for $m=1...M$ with $K_t$ given by Eq.~\eqref{e:Kt};\;\\
calculate the approximate weights for $m=1...M$:
 \[w_{t}^m
= w_{t-1}^m\alpha^m_t,\quad
 \alpha_t^m=\frac{\hat{q}_{t-1}(x^m_t)\pi(y_{t}|\-x^m_t) \hat{L}_{t-1}(x^m_{t-1}|x^m_t)}{\hat{q}_{t-1}(x^m_{t-1})K_{t}(x^m_t|x^m_{t-1})},
\]
and renormalize $\{w_t^m\}_{m=1}^M$ so that $\sum_{m=1}^M w^m_t=1$;\;\\
calculate the ESS of the approximate weights $\{w_t^m\}_{m=1}^M$; \;\\
\If{ESS$<\mathrm{ESS}_{\min}$ $\lor$ $t-t_0>\Delta T_{\max}$ $\lor$ $t=T$}
{
 compute $\hat{L}_{t-1}$ from Eq.~\eqref{e:suboptLnormal};\;\\
calculate the weights for $m=1...M$:
 \[
w^m_t = w^m_{t_0} \frac{\pi_t(x^m_t)}{\pi_{t_0}(x^m_{t_0})}\prod_{i=t_0}^{t-1} \frac{\hat{L}_i(x^m_i|x^m_{i+1})}{K_{i+1}(x^m_{i+1}|x^m_{i})},
\]
and renormalize $\{w_t^m\}_{m=1}^M$ so that $\sum_{m=1}^M w^m_t=1$;\;\\
calculate the ESS of the weights $\{w_t^m\}_{m=1}^M$; \;\\
\If{ESS$<\mathrm{ESS}_\mathrm{resamp}$ }
{resample;}
let $t_0=t$;
}
  }
	\caption{The EnKF-SMCS-WR algorithm}\label{alg:enkf-smcs2}
\end{algorithm}

\section{Numerical examples}\label{sec:examples}
We provide three examples in this section to demonstrate the performance of the proposed method. 
{We emphasize here that in all these examples,  the forward model $G_t$ is computationally intensive and thus 
the main computational cost arises from the simulation of $G_t$. As a result, the main computational cost of all methods is measured
by the number of forward model evaluations, which in all the methods used in this section is equal to the product of the number of steps and 
that of the particles. }

\subsection{The Bernoulli model}

\begin{figure}
  \centering
	\centerline{\includegraphics[width=.5\linewidth]{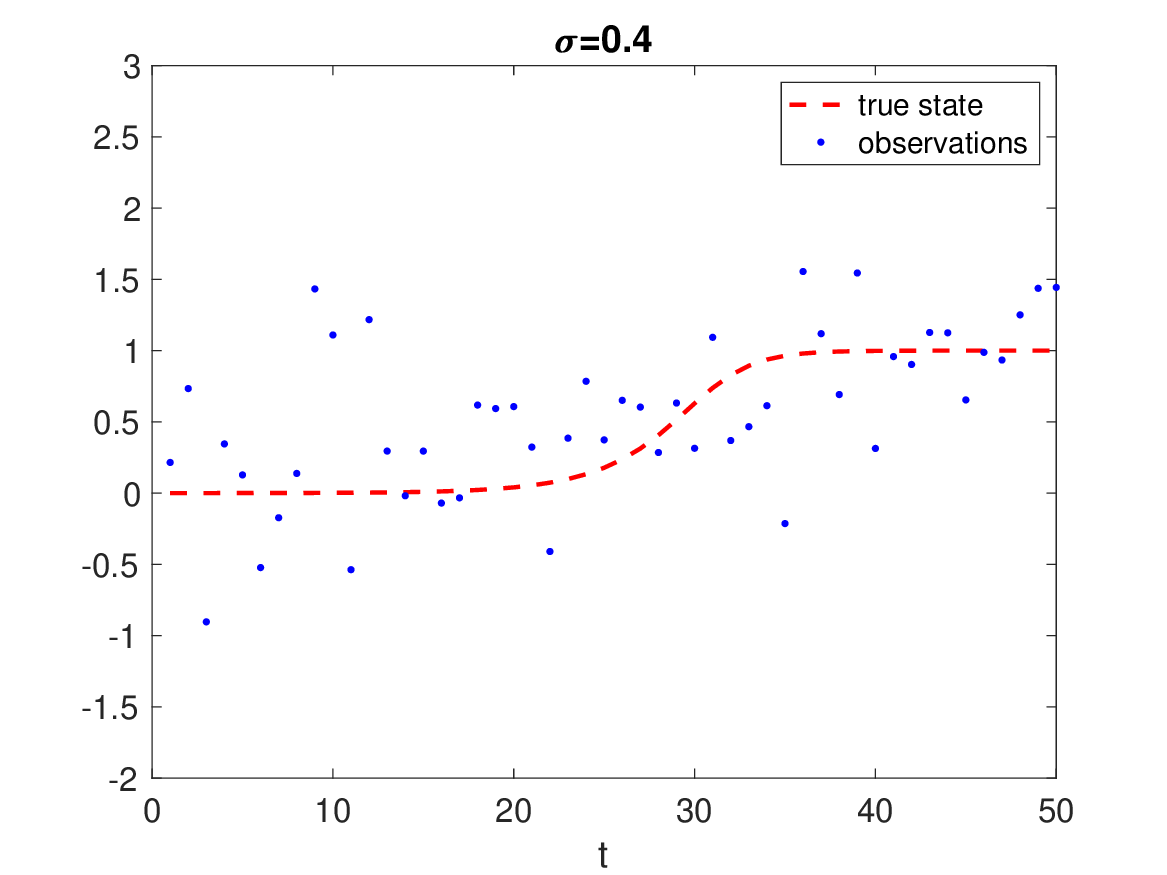}
	\includegraphics[width=.5\linewidth]{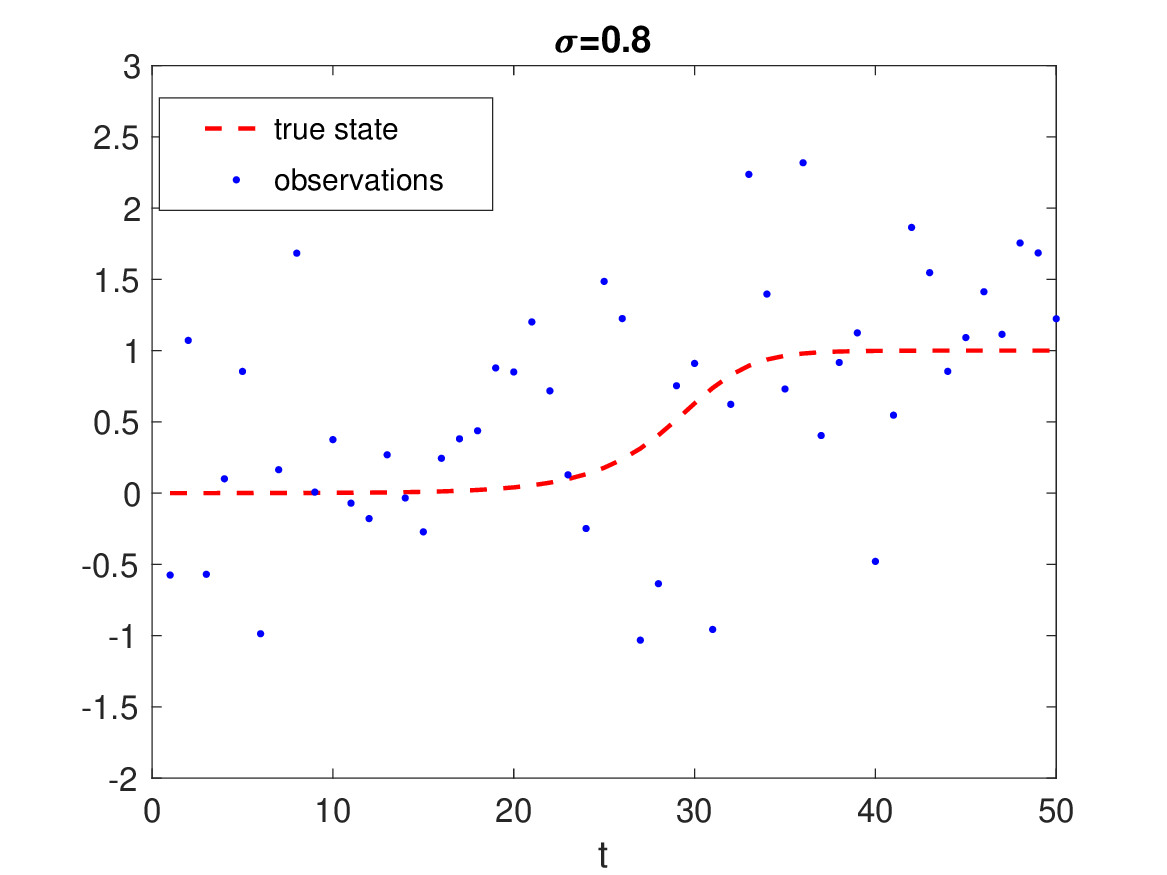}}
	\caption{The simulated data for $\sigma=0.4$ (left) and $\sigma =0.8$ (right).
	The  lines show the simulated states in continuous time  and the dots are the noisy observations. } \label{f:berndata}
\end{figure}
\begin{figure}
  \centering
	\centerline{\includegraphics[width=.5\linewidth]{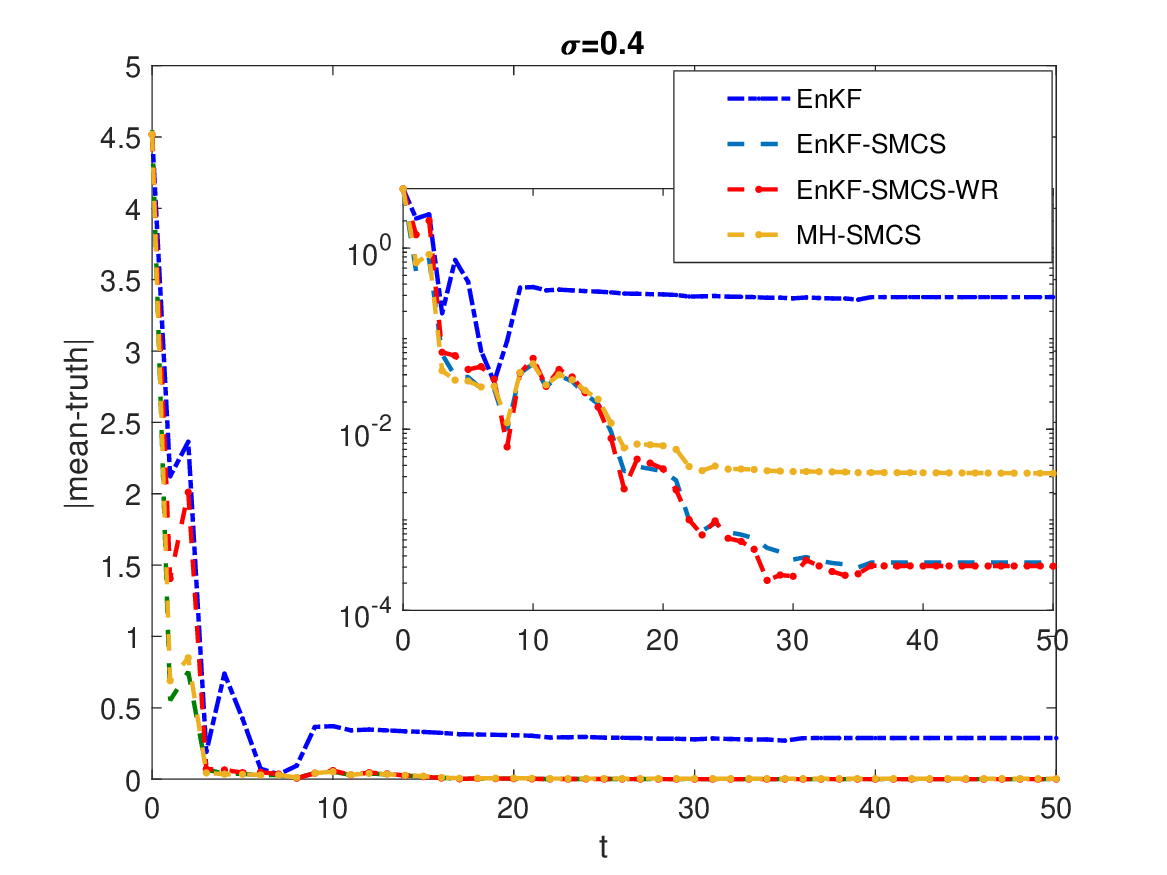}
	\includegraphics[width=.5\linewidth]{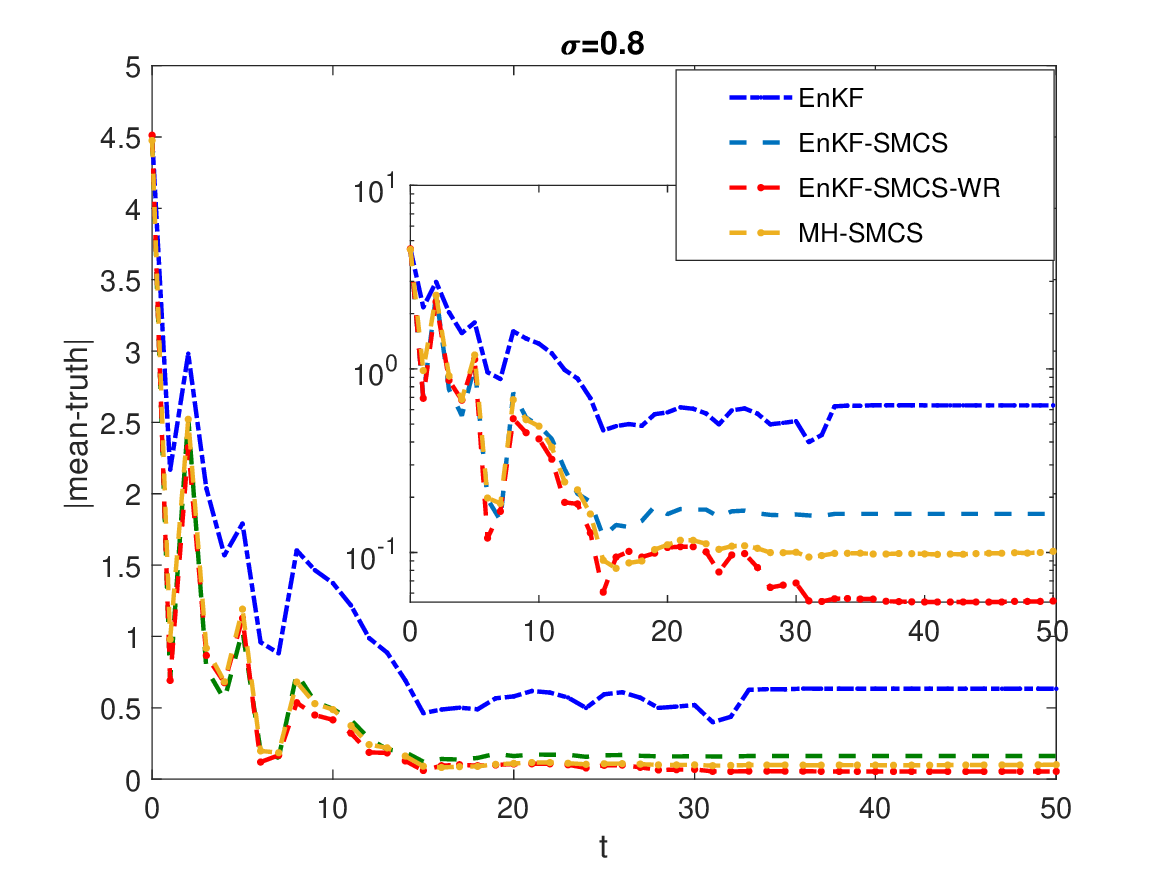}}
	\caption{The average bias error (the difference between the sample mean and the ground truth) plotted at each time step
	where the insets are the same plots on a logarithmic scale. 
	The left plot is the error for $\sigma=0.4$ and the right figure is that for $\sigma=0.8$. } \label{f:bern}
\end{figure}

Our first example is the Bernoulli equation,  
\begin{subequations} \label{e:bern}
	\begin{equation}
	\frac{d v}{d\tau} -v=-v^3,\ \ \ v(0)=x, 
	\end{equation}
	which has an analytical solution, 
		\begin{equation}
	v(\tau)={G}(x,\tau) = x (x^2+(1-x^2)e^{-2\tau})^{-1/2}. \label{e:solution}
	\end{equation}
	\end{subequations}
	This model is an often-used benchmark problem for data assimilation methods as it
	exhibits certain non-Gaussian behavior~\cite{apte2007sampling}. Here we pose it as a sequential inference problem.
	Namely, suppose that we can observe the solution of the equation, $v(\tau)$, at different times $\tau = t\cdot \Delta_t$ 
	for $t=1,...,T$,
	and we aim to estimate the initial condition $x$ from the sequentially observed data.
	The observation noise is assumed to follow a zero-mean Gaussian distribution with standard deviation
	$\sigma$. 
	In this example, we take  $T=50$, and  $\Delta_t=0.3$
and we consider two different noise levels:
$\sigma=0.4$ and $\sigma=0.8$.   
       In the numerical experiments, we set the ground truth to be $x=10^{-4}$ and the data is simulated from the model~\eqref{e:bern} for $\sigma=0.4$ and $\sigma=0.8$, 	which  are shown in Figs.~\ref{f:berndata}.
   In the experiments the prior distribution for $x$ is taken to be uniform: $U[-1,10]$. 

			{We sample the posterior distribution with four methods:   the EnKF method in \cite{iglesias2013ensemble},
			EnKF-SMCS (Algorithm~1), EnKF-SMCS-WR (Algorithm~2) and MH-SMCS.
			Note that MH-SMCS is the SMCS with a Metropolis-Hastings forward proposal, a commonly used 
			implementation of SMCS (we provide a detailed description of the algorithm in  Appendix A).} 
    In each method, we use $200$ particles, and the bias error, i.e., the difference between the sample mean, which is a commonly used estimator, 
    and the ground truth is then computed at each time step. 
    The procedure is repeated 100 times and 
    the averaged results are shown in Figs.~\ref{f:bern} where the left figure show the results for the small noise case ($\sigma=0.4$)
    and the right figure shows those for the large noise case ($\sigma=0.8$).
    First, one can see from the figures that all the methods perform better in the small noise case, which is sensible as intuitively the inference should be more accurate when the observation noise is small. More importantly, we can also see that in both cases, the EnKF results in significantly  higher errors than 
    the three SMCS methods, suggesting that EnKF performs poorly for this example.  
On the other hand, we observe that the three SMCS algorithms produce largely  the same results in both cases, while EnKF-SMCS-WR only calculates the actual sample weights at 9 time steps on average in the small noise case 
    and 6  in the large noise case, as is compared to 50 in the EnKF-SMCS. 
    Such a difference suggests that the  EnKF-SMCS-WR algorithm can significantly 
    reduce the computational cost associated with the weight computation. 
    {The two EnKF-SMCS algorithms and MH-SMCS yield similar results, but we need to emphasize  that 
    the MH-SMCS is substantially more expensive than EnKF-SMCS-WR, as its procedure is similar to that of MCMC (see Appendix \ref{sec:mhsmcs} for details).}

\subsection{Lorenz 63 model}
\begin{figure}
	\centerline{\includegraphics[width=1.25\linewidth]{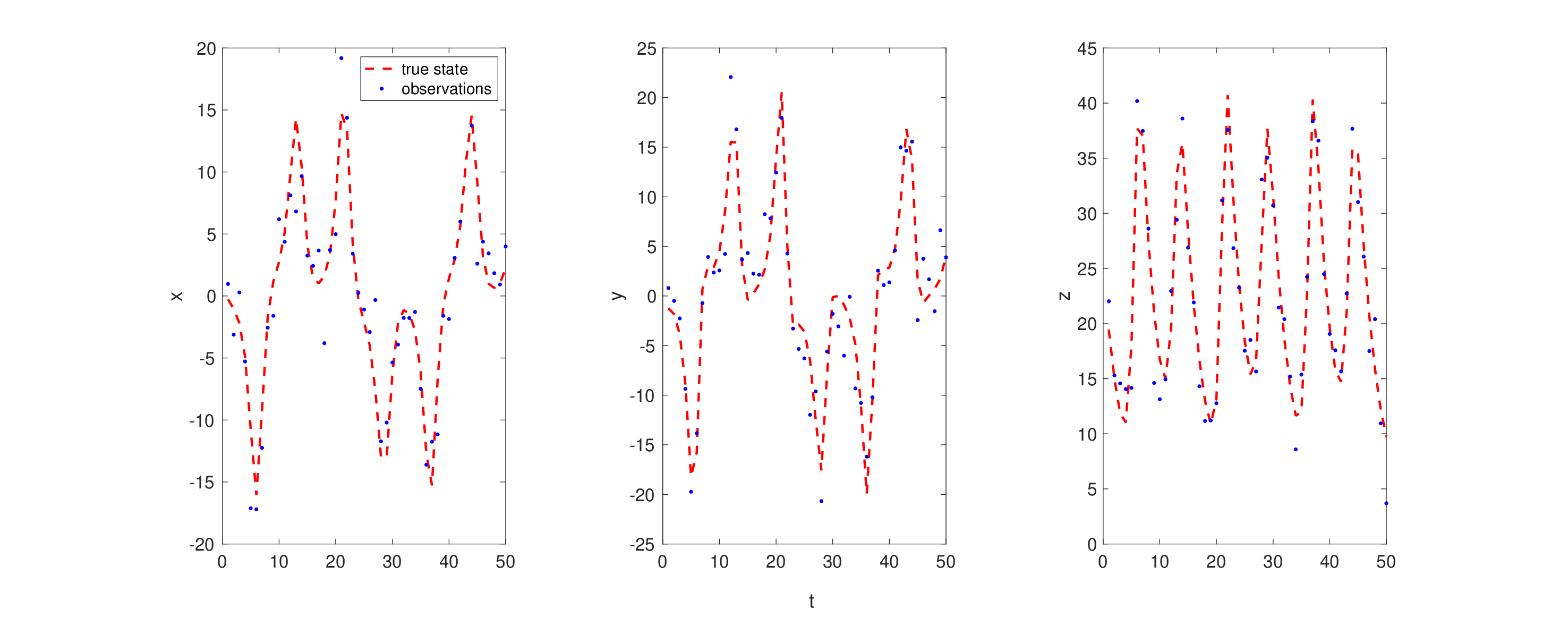}}
	\caption{The simulated data for the Lorenz 63 example. The  lines show the simulated states in continuous time and the dots are the noisy observations.} \label{f:lorenz63data01}
\end{figure}
\begin{figure}
	\centerline{\includegraphics[width=1\linewidth]{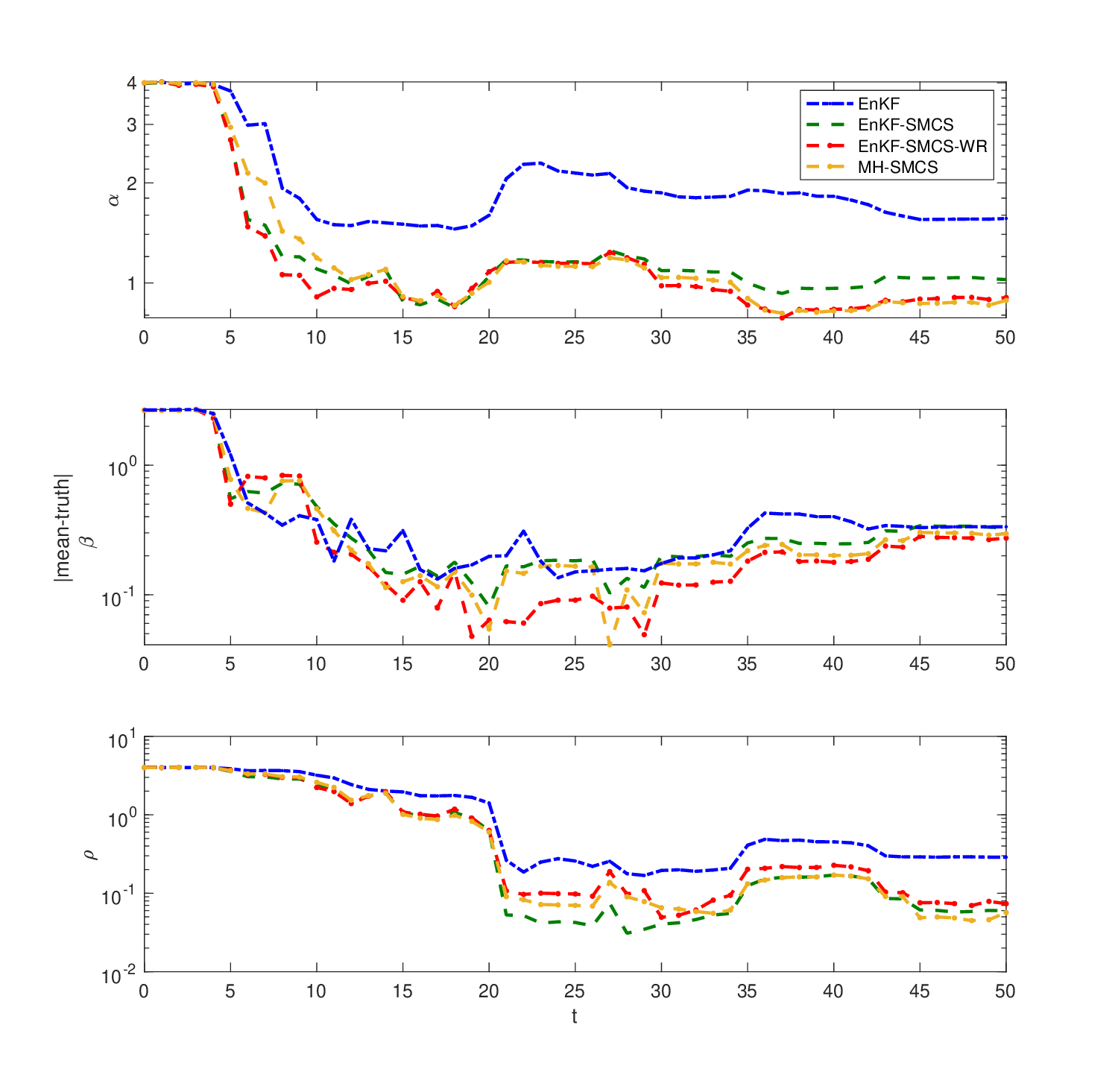}}
	\caption{The average estimation error of each parameter when $x$ is observed.}\label{f:x1}
	\end{figure}
	\begin{figure}
	\centerline{\includegraphics[width=1\linewidth]{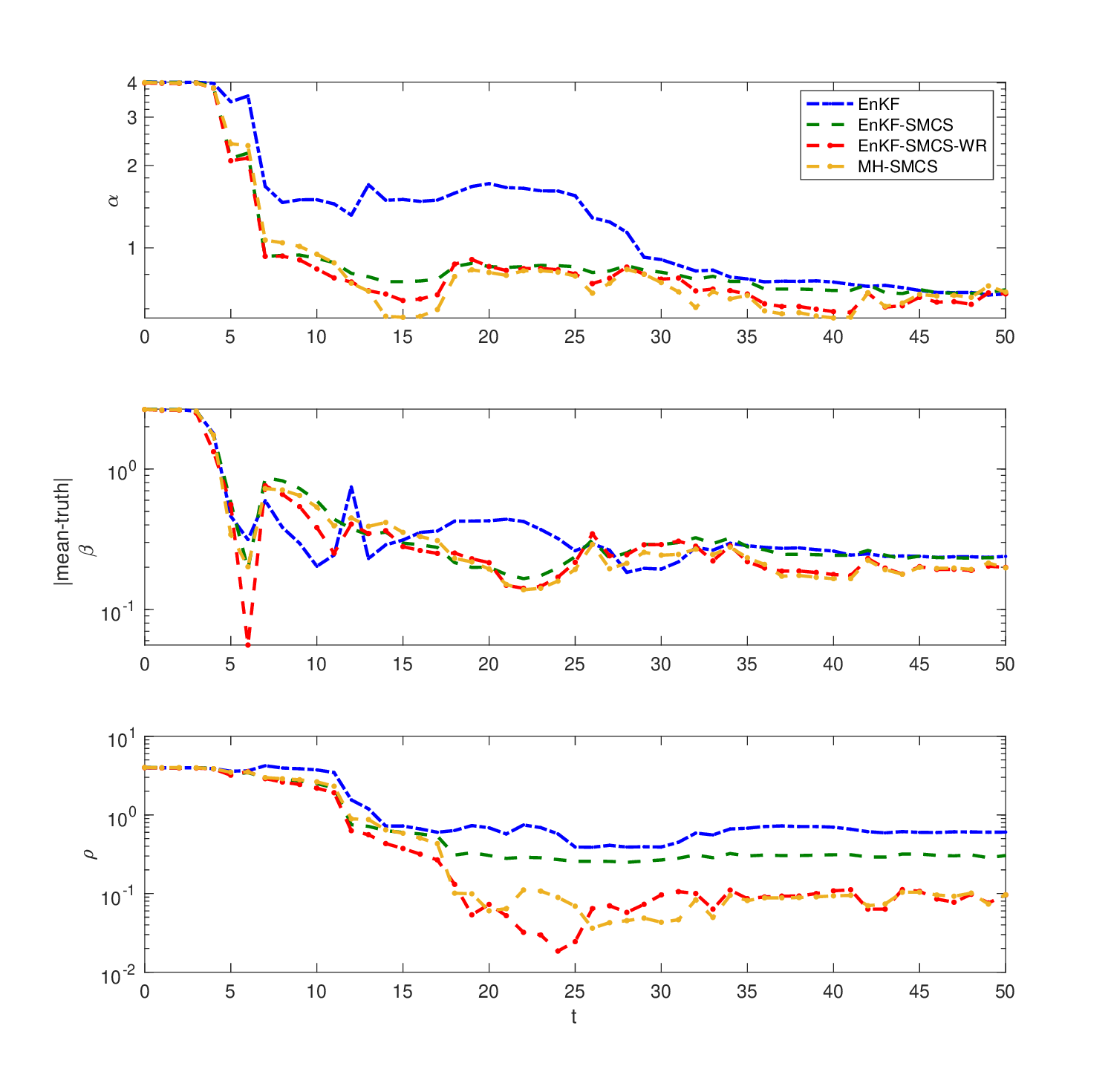}}
	\caption{The average  estimation  error of each parameter when $y$ is observed. } \label{f:x2}
\end{figure}


Our second example is the Lorenz 63 model, a popular example used in several works on parameter estimation, such as 
\cite{annan2004efficient,mehrkanoon2012parameter}. 
Specifically the model consists of three
variables $x$, $y$ and $z$, evolving according to the differential equations
\begin{subequations}
\label{e:lorenz63}
\begin{eqnarray}
\frac{dx}{d\tau}&=& \alpha(y-x),\\
\frac{dy}{d\tau}&=& x(\rho -z)-y,\\
\frac{dz}{d\tau}&=& xy-\beta z,
\end{eqnarray}
\end{subequations}  
where $\alpha$, $\rho$ and $\beta$ are three constant parameters.
In this example we take the true values of the three parameters to be
$\alpha=10$, $\beta=8/3$ and $\rho=28$, which we assume that we have no knowledge of. 
Now suppose that  observations of $(x,y,z)$ are made at a sequence of discrete time points: $\tau = t\cdot \Delta_t$  
	for $\Delta_t=0.1$ and $t=1,...,50$,  
and we want to estimate the three parameters $(\alpha,\beta,\rho)$ from these observed data.
The measurement noise here is taken to be zero-mean Gaussian with variance $3^2$, and the priors of the three parameters are also taken to be  Gaussian with means {$[6, 0, 24]$}, and variances $[1, 1 ,1]${ (the prior is chosen so that it covers the regime that can result in chaotic behavior) }. 
The data used in our numerical experiments are shown in Fig.~\ref{f:lorenz63data01}.

	\begin{figure}
	\centerline{\includegraphics[width=.75\linewidth]{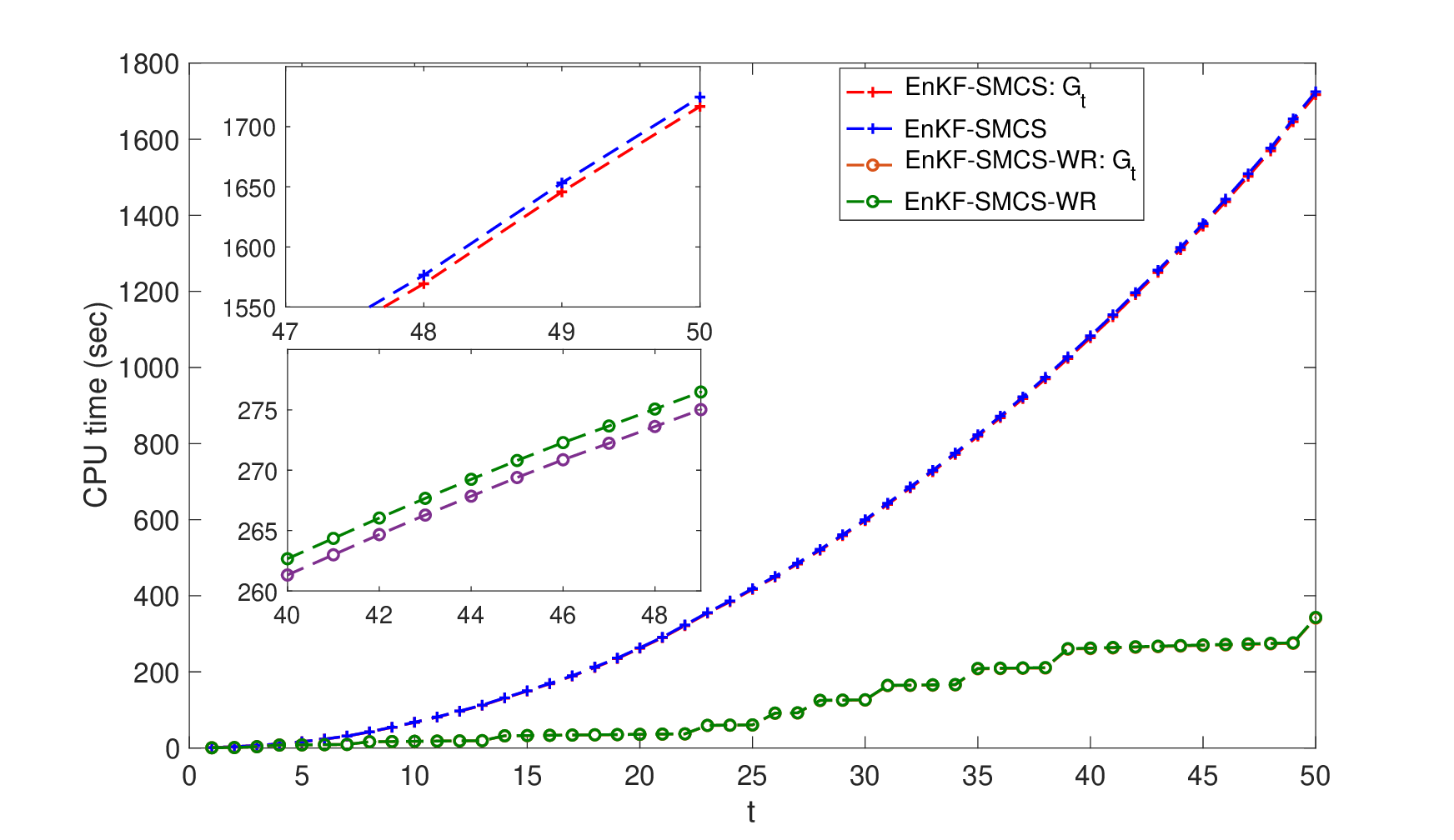}}
	\caption{The total CPU time and the CPU time for the forward model evaluation (marked with $G_t$) in both EnKF-SMCS and EnKF-SMCS-WR.
	Insets are the zoom-in plots.}\label{f:cputime}
	\end{figure}

	\begin{figure}
	\centerline{\includegraphics[width=.75\linewidth]{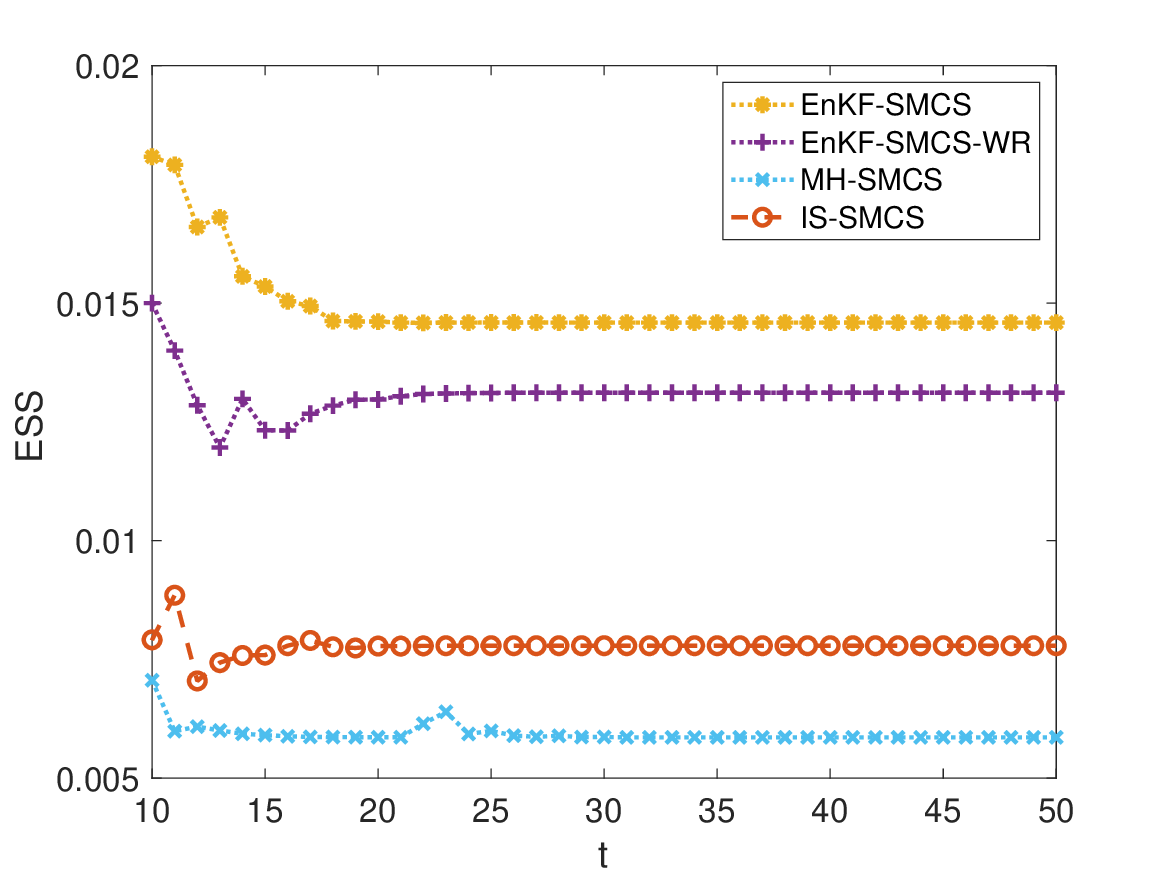}}
	\caption{The ESS (without resampling) plotted as a function of $t$.}\label{f:ess}
	\end{figure}	
	
In the numerical experiments, we conduct inference for two different cases:  one is that variable $x$ is observed and the other is that $y$ is observed.  In each case we draw samples from the posterior distributions with EnKF, EnKF-SMCS and EnKF-SMCS-WR, MH-SMCS, where 
$500$ samples are drawn with each method. All the numerical experiments are repeated 10 times. 
We plot the average errors for the case that $x$ is observed in Fig.~\ref{f:x1}
and those for that with $y$ being observed in Fig.~\ref{f:x2}. 
One can see that, in both cases, the errors in the EnKF is larger than those in the three SMCS methods, especially 
for parameter $\alpha$. Once again, the three SMCS methods yield similar errors while 
EnKF-SMCS-WR employs much less computations of the actual weight:  on average 9 time steps in the first case and 8 in the second. 
The example shows that even for problems where the posterior distributions are rather close to Gaussian, the SMCS can further 
improve the estimation accuracy. 

\subsection{A kinetic model of the ERK pathway}
In the last example we consider the parameter estimation problems in the kinetic models. 
Estimating the kinetic parameters is an essential task in the modeling of the biochemical reaction networks, including
genetic regulatory networks and signal transduction pathways~\cite{quach2007estimating}.
In particular we consider the kinetic model of the Extracellular signal Regulated Kinase (ERK) pathway suppressed by Raf-1 kinase inhibitor protein (RKIP)~\cite{kwang2003mathematical,sun2008extended}.
Here we shall omit further details of biological background of the problem and proceed directly to the mathematical formulation of the problem;
 readers who are interested in more application-related information may consult \cite{kwang2003mathematical,sun2008extended}.

In this problem the mathematical model that is derived based on enzyme kinetics, 
and is represented by a dynamical system:
	\begin{equation}
 \frac{dx}{d\tau} = SV(x),  \label{e:kinetic}
	\end{equation}
where $\tau$ is the time, $x$ is a vector of state
variables which are concentrations of metabolites, enzyme and
proteins or gene expression levels,  $S$ is a stoichiometric matrix that describes the biochemical
transformation in a biochemical network, and
$V(x)$ is the vector of reaction rates and is usually the vector of
nonlinear function of the state and input variables. Specifically, in this ERK pathway model we have
$$
x=[x_1, x_2,...,x_{11}]^T,\quad V(x)=[v_1, v_2,...,v_7]^T,$$
which forms a system of 11 ordinary differential equations. 
Moreover the rates of reactions $V(x)$ are~\cite{kwang2003mathematical,sun2008extended}:
\begin{subequations} \label{e:kinetic2}
	\begin{align*}
	& v_1 = k_1x_1x_2-k_2x_3,  \quad
	 v_2 = k_3x_3x_9-k_4x_4, \\
       & v_3 = k_5x_4,  \quad
	 v_4 = k_6x_5x_7-k_7x_8,  
	\\ & v_5 = k_8x_8,  \quad
	 v_6 = k_9x_6x_{10}-k_{10}x_{11},  \quad
	 v_7 = k_{11}x_{11},
	\end{align*}
	\end{subequations}
	where $k_1,...,k_{11}$ are the kinetic parameters, 
and the stoichiometric matrix $S$ is given by~\cite{kwang2003mathematical,sun2008extended}:	
	\begin{align*}
 S =& \begin{bmatrix}
   -1&  0&  1&  0&  0&  0&  0\\
   -1&  0&  0&  0&  0&  0&  1\\
    1& -1&  0&  0&  0&  0&  0\\
    0&  1& -1&  0&  0&  0&  0\\
    0&  0&  1& -1&  0&  0&  0\\
    0&  0&  1&  0&  0& -1&  0\\
    0&  0&  0& -1&  1&  0&  0\\
    0&  0&  0&  1& -1&  0&  0\\
    0& -1&  0&  0&  1&  0&  0\\
    0&  0&  0&  0&  0& -1&  1\\
    0&  0&  0&  0&  0&  1& -1
\end{bmatrix}.
	\end{align*}

In this problem,  we can make observations of some of the concentrations $x_1,...x_{11}$ at different times,
from which we estimate the 11 kinetic parameters $k_1,...,k_{11}$. 
In our numerical experiments the specific setup is the following.
In many practical problems, not all the species' concentrations can be conveniently observed \cite{kwang2003mathematical,sun2008extended}. To mimic the situation we assume that the observations can only be made on 4 of the states:
 $\{x_1, x_4, x_7, x_{10}\}$, and $\{x_2, x_3, x_5, x_6, x_8, x_9, x_{11}\}$
		are  not observed.
Second the observation is made 50 times with each time spacing is $\Delta_t=0.001$,
and the measurement noise is taken to be zero-mean Gaussian with standard deviation (STD) shown in Table~\ref{tab:states}.
The initial values of the concentrations are also given in Table~\ref{tab:states}.
We use simulated data in this example where 
the true values of the eleven parameters are shown in Table~\ref{tab:paras}. 
The prior of the eleven parameters are also taken to be Gaussian with means and standard deviations both shown in Table~\ref{tab:paras}.

In this example we focus on the four SMCS algorithms: EnKF-SMC, EnKF-SMC-WR, MH-SMCS, and IS-SMCS (a special MH-SMCS implementation with an independent proposal;
see  Appendix~\ref{sec:mhsmcs} for details),
where the main purpose is to compare the EnKF based and the MH based forward proposals. 
We test two sample sizes $M=5000$ and $M=10000$ for each method 
and all the tests are repeated 10 times.

{First we want to examine the computational cost of EnKF-SMCS and EnKF-SMCS-WR. 
To do this, we plot the CPU time of the two algorithms as a function of $t$, where in each algorithm we show both the total time cost and that used for evaluating the forward model. 
First, we can see from the figure that, in both algorithms the main computational cost arises from the forward model evaluation;
second, the EnKF-SMCS-WR can significantly reduce the computational cost by using less forward model evaluations.}
As discussed earlier, for the purpose of sequential inference, we should devote the majority of our attention to the estimator accuracy at the final step, 
and therefore in Table~\ref{tab:11d2_error} we show 
the estimation error for each parameter at the final step $t=50$.
Specifically, we provide in the table the mean-squared errors (MSE) of the estimation results. 
We can see from the table that the two EnKF-SMCS algorithms yield lower estimation errors 
than MH-SMCS in all the cases,
and in particular the difference is substantially large for parameters $k_1$, $k_4$, $k_5$ $k_6$, $k_8$, $k_9$ and $k_{11}$,
with both sample sizes. 
{The results of IS-SMCS are considerably better than those of MH-SMCS, 
suggesting that the posterior distributions in this problem may be reasonably close to Gaussian.
That said, IS-SMCS results in clearly higher MSE for $k_1$ and $k_9$ (in the $M=10000$ case). }
On the other hand, the two EnKF based SMCS algorithms yield similar performance in terms of the estimation error,
but the EnKF-SMCS-WR method only conducts WR at 12 steps on average for both sample sizes, 
resulting in much higher computational efficiency than EnKF-SMCS.
%

\begin{table}[htbp]
\caption{The initial values and observation noise of the concentrations (states $x_i$)}
\label{tab:states}
\centering
\begin{tabular}{c|cccccc} \hline
		& $x_1$&$x_2$&$x_3$&$x_4$&$x_5$&$x_6$\\ \hline  
  initial\ values& 66& 0.054& 0.019& 59& 0.09& 0.012  \\
  noise STD &0.005& $5\times 10^{-5}$& $2\times 10^{-5}$&  0.035& 0.0005& $5\times 10^{-6}$ \\  \midrule     
  &$x_7$&$x_8$&$x_9$&$x_{10}$&$x_{11}$&  \\ \hline
  initial\ values& 65& 26& 175& 161& 2.18&       \\
  noise STD & 0.05& 0.02& 0.03& 0.003& 0.002&      \\  \hline
\end{tabular}
\end{table}

\begin{table}[htbp]
\caption{The true values and priors of the kinetic parameters}
\label{tab:paras}
\centering
\begin{tabular}{c|cccccc} \hline
		& $k_1$&$k_2$&$k_3$&$k_4$&$k_5$&$k_6$  \\ \hline				
  truth &0.5242& 0.0075& 0.6108& 0.0025& 0.0371& 0.8101  \\ \hline
  prior mean&0.5&  0.1&   0.62&   0.04&   -0.5&  0.8  \\
  prior STD&0.05&   0.03&  0.01&  0.04&     0.5&  0.02  \\ \midrule  
  
  &$k_7$&$k_8$&$k_9$&$k_{10}$&$k_{11}$& \\ \hline  
  truth & 0.0713& 0.0687& 0.96& 0.0012& 0.872&   \\  \hline  
  prior mean&    0&        0.4&  0.9&   0&          0.9&     \\
  prior STD&  0.05&  0.3&   0.1&   0.005&  0.05&          \\  \hline  
\end{tabular}
\end{table}	

\begin{table}[htbp]

\caption{ Comparison of the MSE results of the kinetic model at  t=50.}

\label{tab:11d2_error}
\centering
\begin{tabular}{c|cccc}  \hline  
 & & $\bf{M\,= 5000}$ && 


\\ \hline

&{$\bf{EnKF}$}&{$\bf{EnKF}$}&{$\bf{MH}$ }&{$\bf{IS}$ }\\


  &{$\bf{SMCS-WR}$}&{$\bf{SMCS}$}&{$\bf{SMCS}$}&{$\bf{SMCS}$}  \\ \hline

{$k_1$}
&0.0078&0.0094& 0.0287&0.0215 
\\ \hline

{$k_2$}
&0.0877&0.0915 &  0.1011&0.0920 
\\ \hline

{$k_3$}
&0.0093&0.0091&	0.0119&0.0094  
 \\ \hline

{$k_4$}
&0.0080&0.0078 &	0.0201&0.0075  
\\ \hline

{$k_5$}
&0.0003&0.0002& 	0.0011&0.0001 
\\ \hline

{$k_6$}
&0.0091&0.0092& 	0.0202  &0.0100 
\\ \hline

{$k_7$}
&0.0668&0.0651& 	0.0936 &0.0669 
\\ \hline

{$k_8$}

&0.0199&0.0191&  	0.0421&0.0307 
\\ \hline

{$k_9$}

&0.0108&0.0065& 	0.0888& 0.0111 
\\ \hline

{$k_{10}$}

&0.0012&0.0011& 	0.0016&0.0015 
\\ \hline

{$k_{11}$}

&0.0068&0.0056&	0.0299 & 0.0045 
\\ \hline\hline
&&$\bf{M\,=10000}$ 
\\\hline
&{$\bf{EnKF}$}&{$\bf{EnKF}$}&{$\bf{MH}$ }&{$\bf{IS}$ }\\


  &{$\bf{SMCS-WR}$}&{$\bf{SMCS}$}&{$\bf{SMCS}$}&{$\bf{SMCS}$}  \\ \hline
  {$k_1$}


 &0.0092&0.0082 & 0.0429 & 0.0213 
\\ \hline

{$k_2$}

 

&0.0906&0.0928& 0.0942 & 0.0922 
\\ \hline

{$k_3$}


&0.0095 & 0.0101&0.0113 &0.0092 \\ \hline

{$k_4$}

  
& 	0.0080&0.0082& 0.0128 & 0.0071
\\ \hline

{$k_5$}


&0.0001&0.0002&	0.0027 &0.0001
\\ \hline

{$k_6$}


&	0.0101& 0.0097& 0.0177 & 0.0106
\\ \hline

{$k_7$}


&0.0695&0.0687 & 	0.0946& 0.0653
\\ \hline

{$k_8$}


& 	0.0207&0.0206&  	0.0373 & 0.0289 
\\ \hline

{$k_9$}


& 	0.0023&0.0040 &	0.0532 & 0.0125
\\ \hline

{$k_{10}$}


& 	0.0014&0.0015 &  	0.0028& 0.0014
\\ \hline

{$k_{11}$}


& 	0.0023&0.0036& 	0.0652 & 0.0058 
\\ \hline

\end{tabular}

\end{table}

\section{Conclusions}\label{sec:conclusions}
In this work we propose a sampling method to compute the posterior distribution that arises in sequential Bayesian inference problems. 
The method is based on SMCS, which seeks to generate weighted samples from the posterior in a sequential manner
and, specifically, we propose to construct the forward kernel in SMCS using 
an EnKF framework and also derive a backward kernel associated to it.  
With numerical examples, we demonstrate that the EnKF-SMCS method can often yield more accurate estimations
than the direct use of either SMCS or EnKF for a class of problems. We believe that the  method can be useful in
a large range of real-world parameter estimation problems where  data becomes available sequentially in time. 

Some extensions and improvements of  the EnKF-SMCS algorithm are possible. First in this work we focus on problems with a sequential structure,
but we expect that the method can be applied to batch-inference problems (where the data are available and used for inference altogether) as well. 
In fact, many batch inference problems can be artificially ``sequentialized'' by some
data tempering treatments~\cite{geyer2011importance} and, consequently, the EnKF-SMCS algorithm can be applied in these scenarios. 
In this respect, combining data tempering methods and the EnKF-SMCS method to address batch inference problems 
can be a highly interesting research problem. 
Second, as has been discussed previously, the proposed method relies on the assumption 
that the posterior distributions do not deviate strongly from being Gaussian. 
For problems with highly nonlinear models, the posterior distributions may depart far from Gaussian, and as result the kernels obtained with the EnKF method may not be effective for SMCS. 
In this case, the performance of the EnKF-SMCS method may be improved by approximating the 
posterior with a mixture distribution (e.g. \cite{hoteit2008new,stordal2011bridging}). 
{Finally, as the method is based on an EnKF scheme, it requires that the observation noise is additive Gaussian. 
In the EnKF literature, a number of methods have been developed to deal with non-Gaussian observations. 
One simple approach is to calculate the Kalman gain matrix using the sample covariance  between $u_t$ and $y_t$~\cite{houtekamer2001sequential},
and another example is the EnKF variant proposed in \cite{lei2011moment}. 
In principle, all these ideas  can be used in our method to construct the EnKF forward kernel,
and to this end an important question is how to effectively incorporate them in the EnKF-SMCS framework.}
We plan to investigate these issues in the future.

\appendix
\section{The MH-SMCS algorithms} \label{sec:mhsmcs}
In this section we provide a brief description of the MH-SMCS algorithm, largely following \cite{del2006sequential}. 
As is mentioned  in section 6.1, a commonly used forward proposal is the MH kernel, which generates a new sample $x_t$ 
with the following procedure, 
\begin{itemize}
\item Draw $x^*$ from a condition distribution $q(\cdot|x_{t-1})$; 
\item Calculate the acceptance probability $a(x^*,x_{t-1})= \min\{1,\frac{\pi_t(x^*)}{\pi_t(x_{t-1})} \frac{q(x_{t-1}|x^*)}{q(x^*|x_{t-1})}\}$;
\item Draw $\nu$ from $U[0,1]$;
\item If $\nu\leq a$, let $x_t=x^*$; otherwise let $x_t=x_{t-1}$.
\end{itemize}
We then choose the backward kernel given by Eq. (32)  in  \cite{del2006sequential}, and the resulting incremental weight  is 
\[ \alpha_t =\frac{\pi_t(x_{t-1})}{\pi_{t-1}(x_{t-1})}.\]
It is worth mentioning that, since the MH kernel includes an acceptance-rejection step that involves  $\pi_t(\cdot)$, 
the number of forward model evaluations is also at the order of $O(T^2)$ where $T$ is the total number of time steps.
Finally we note that, in the numerical experiments, 
we choose two proposal distribution within the MH kernel: (1) a random walk proposal, $q(\cdot|x_{t-1})=\@N(x_{t-1},\lambda^2 I)$,
and the variance $\lambda^2$ is adjusted so that the average acceptance rate is about 50\%;
(2) an independence sampler proposal $q(\cdot|x_{t-1})=\hat{q}_{t-1}(\cdot)$ where $\hat{q}_{t-1}(\cdot)$ is given
by Eq.~\eqref{e:qhat}.
We refer to the former as MH-SMCS and the latter as IS (independent sampler)-SMCS.
Finally we note that, in principle multiple-iteration MH proposal can be used in SMCS; however, since each MH iteration requires an evaluation of the forward model, 
which may result in formidable computational cost,  here we only perform one MH iteration in both MH-SMCS and IS-SMCS. 
    
\bibliographystyle{plain}
\bibliography{smcs}
\end{document}